\documentclass[conference]{IEEEtran}
\IEEEoverridecommandlockouts
\usepackage{cite}
\usepackage{amsmath,amssymb,amsfonts}
\usepackage{algorithmic}
\usepackage{graphicx}
\usepackage{textcomp}
\usepackage{xcolor}
\def\BibTeX{{\rm B\kern-.05em{\sc i\kern-.025em b}\kern-.08em
    T\kern-.1667em\lower.7ex\hbox{E}\kern-.125emX}}
\begin{document}

\title{Opportunities and challenges of Blockchain-Oriented systems in the tourism industry. }

 \author{
 \IEEEauthorblockN{Fabio Caddeo
 }
 \IEEEauthorblockA{
 %\textit{Department of Economics and Business Sciences} \\
 \textit{University of Cagliari}\\
 Cagliari, Italy \\
 fabio.caddeo@yahoo.it}

 \and
 \IEEEauthorblockN{Andrea Pinna}
 \IEEEauthorblockA{\textit{Department of Mathematics and Computer Science } \\
 \textit{University of Cagliari}\\
 Cagliari, Italy \\
 pinna.andrea@unica.it}
}

\maketitle

\begin{abstract}

%L’industria del turismo è sempre più influenzata dall’evoluzione delle tecnologie dell’informazione e della comunicazione (ICT), che stanno rivoluzionando il modo di viaggiare delle persone.
%In questo lavoro si vuole indagare l’utilizzo delle tecnologie informatiche innovative da parte delle DMO (Destination Management Organization), concentrando l’attenzione sulla tecnologia blockchain, sia dal punto di vista della ricerca nel settore, che nello studio dei progetti software più rilevanti. In particolare, si intende verificare i benefici offerti da questi strumenti informatici nella gestione e nel monitoraggio di una destinazione, senza dimenticare le implicazioni per gli altri stakeholder coinvolti. Tali tecnologie, infatti, possono offrire un vasto insieme di servizi che possono risultare utili durante l’intero ciclo di vita della destinazione. 

The tourism industry is increasingly influenced by the evolution of information and communication technologies (ICT), which are revolutionizing the way people travel.
In this work we want to investigate the use of innovative IT technologies by DMOs (Destination Management Organizations), focusing on blockchain technology, both from the point of view of research in the field, and in the study of the most relevant software projects. In particular, we intend to verify the benefits offered by these IT tools in the management and monitoring of a destination, without forgetting the implications for the other stakeholders involved. These technologies, in fact, can offer a wide range of services that can be useful throughout the life cycle of the destination.

\end{abstract}

\begin{IEEEkeywords}
Blockchain, BOSE, Tourism industry, E-Tourism
\end{IEEEkeywords}

\section{Introduction}

Nowadays, tourists are increasingly involved in creating the travel experience, becoming promoters of what they experience first-hand and consequently contributing to the improvement of a destination. In other words, they become protagonists of a process of co-creation of value that can affect both the destination in general and individual companies operating in the sector; examples are "Expedia", one of the largest worldwide Online Travel Agencies (or OTA), and "TripAdvisor", a travel platform that offers the opportunity of reviewing the tourist services that you have enjoyed on holiday (not just accommodations, restaurants, or airlines, but also experiences and attractions). In fact, through the involvement of their customers / travellers, they have managed to increase their importance in the travel industry. 
With the spread of the Internet, the world of tourism has undergone a first transformation, leading to the spread of the so-called e-tourism. More precisely, according to the definition provided by Buhalis (2003)\cite{buhalis2003}, e-tourism means "the digitization of all processes and value chains in the tourism, travel, hospitality and catering sectors that allow organizations to maximize their efficiency and effectiveness ". This situation has given a strong impetus to the tourism sector, as it has allowed the various destinations to exploit this technology to improve their attractiveness by implementing appropriate online communication strategies. Recently, however, we are witnessing the transition from e-tourism to smart tourism.

%In realtà, però, piuttosto che di smart tourism sarebbe più giusto parlare di smart destination, che Lopez de Avila (2015) definisce come: “una destinazione turistica innovativa, costruita su un’infrastruttura di tecnologie all’avanguardia, che garantisce lo sviluppo sostenibile di aree turistiche, accessibile a tutti, e che facilita l’integrazione del visitatore con il suo ambiente, aumentando la qualità dell’esperienza e migliorando la qualità della vita dei residenti”. Da tale definizione emerge chiaramente l’importanza delle ICT ai fini della loro integrazione nell’infrastruttura fisica delle destinazioni.

In truth, however, rather than smart tourism it would be more right to talk about smart destinations, which Lopez de Avila (2015)\cite{lopez2015} defines as: "an innovative tourist destination built on a state-of-the-art technology infrastructure, which ensures the sustainable development of tourist areas, accessible to all, and which facilitates the visitor's integration with its environment, increasing the quality of the experience and improving the quality of life of residents”. This definition clearly shows the importance of ICTs for their integration into the physical infrastructure of the destinations.

%Il ricorso alle tecnologie dell’informazione è quindi un requisito essenziale delle nuove forme di turismo poiché, da un lato, queste aiutano i turisti nell’identificazione e acquisto dei prodotti turistici che più preferiscono, rendendoli sempre più coinvolti nella co-creazione e co-promozione delle esperienze di viaggio, e dall’altro consentono ai fornitori di tali prodotti di far conoscere le loro offerte in tutto il mondo.

The use of information technologies is therefore an essential requirement of new forms of tourism. On the one hand, they help tourists in identifying and purchasing the tourist products they prefer, making them increasingly involved in co-creation and co-promotion of travel experiences, and on the other hand they allow suppliers of such products to make their offerings known around the world.

One of the problems arising from the use of new technologies in tourism is the lack of public confidence. Blockchain technology would seem to provide a solution, as the strengthening of trust represents one of the potential effects deriving from its application to tourism, especially in a period of great uncertainty like the one we are currently experiencing. The formation of trust in the hospitality industry, in fact, is still a little-known aspect, since it depends on the subjectivity of individual tourists, or rather on the risk that they are willing to accept in travel experiences. In the specific case of blockchain technology, the protocols on which it is based are structured in such a way as to ensure greater involvement by the various tourism stakeholders and thus improve their experience in the sector.

As highlighted by Porru et al. (2017)\cite{porru2017}, the growing focus on the world of blockchain technology has led to the need to create tools for the development of specific software oriented towards it (called blockchain-oriented software, or BOS). In general, BOS is defined as a system that works through the implementation of a blockchain. An example is the Ethereum platform, which can be defined as the largest decentralized digital platform in the world that uses blockchain technology for the realization not only of transactions, but also of particular programs called smart contracts, ensuring security, reliability and transparency\cite{pierro2020,fenu2018}. More precisely, "Smart Contracts" are programs written in a programming language and registered in the blockchain that self-execute when certain predetermined conditions occur. Pinna et al. (2019)\cite{pinna2019} define Smart Contracts as programs stored within the public register of Ethereum and associated with a particular blockchain address, aimed at implementing a logical sequence of steps according to some well-defined rules. 

In this work we want to investigate the use of innovative IT technologies by DMOs, focusing attention on blockchain technology.  In particular, we intend to verify the potential offered by these tools in the management and monitoring of a destination, without forgetting the implications for the other stakeholders involved. Such technologies, in fact, can offer a wide range of services that can be useful throughout the entire destination life-cycle. They represent an indispensable tool both to provide DMOs with different advantages from the management point of view (for example in monitoring or marketing) in order to achieve the sustainability objective\cite{pinna2021}, both to ensure that tourists and operators in the sector have a satisfactory experience. 

The investigation discussed in this paper is guided by the following research questions.

RQ1: to what extent does the scientific community address the problems relating to the use of blockchain technology in tourism? and

RQ2: what is the state of the art of the practical use of blockchain technology in tourism and DMOs?

To answer the first question, we examined the scientific literature on the SCOPUS database, and we deepened the study by examining the content of the works more related to technical and socio-economic issues, the results of which are presented in Section \ref{literature}.
To answer the second question, we collected information regarding the most relevant blockchain-oriented projects for tourism, analyzing both the functionalities and the technology used. Section 3 presents the study of eleven software projects currently operational or under development.
Through the study of the scientific literature of reference and documentation on specialized websites, it is intended to firstly evaluate the contribution of these technologies to tourism managers, by laying the foundations for subsequent implementations.
Furthermore, the Section 4 of the paper allows us to discuss the two directions towards which the tourism industry is moving towards thanks to the use of blockchain technology, i.e disintermediation and coordinaion \& coopetition, as is arising from our investigation. Finally, Section 5 concludes the paper.

\section{Literature Review}\label{literature}

%Nonostante la letteratura scientifica sul tema della blockchain sia sempre più ampia, sono ancora pochi i contributi che la analizzano in relazione all’industria turistica, ma tutti si prestano come base per nuovi approfondimenti. A conferma di ciò, nella Fig. \ref{research} si può verificare il numero di articoli che affrontano il tema della blockchain legata al turismo pubblicati annualmente. I dati sono stati recuperati nel mese di novembre 2020 attraverso il ricorso al database Scopus, impostando la ricerca tramite la query $TITLE-ABS-KEY ( "blockchain" \ AND \ "tourism" )$ con la quale sono stati estratti tutte le pubblicazioni scientifiche contenenti le parole “blockchain” e “tourism” nel titolo, nell’abstract oppure tra le parole chiave.  Il grafico mostra come a partire dal 2016, con nessun articolo pubblicato su questo tema, il numero sia raddoppiato di anno in anno, il che evidenzia il crescente interesse della comunità scientifica verso l’applicazione della tecnologia blockchain nel settore turistico.

Although the scientific literature on the topic of blockchain is increasingly extensive, there are still few contributions that analyse it in relation to the tourist industry, but all are suitable for new insights. To confirm this, in Fig. \ref{research} you can check the number of articles published (or added to the lists of papers in press) annually dealing with the issue of blockchain related to tourism. The data were recovered in November 2020 through the use of the Scopus database, setting the search by query $TITLE-ABS-KEY ( "blockchain" \ AND \ "tourism" )$ with which were extracted scientific papers containing the words “blockchain” and “tourism” in the title, in the abstract or among the keywords. The graph shows that since 2016, with no article published on this theme, the number has doubled from year to year, which highlights the growing interest of the scientific community towards the application of blockchain technology in the tourism sector.

\begin{figure}[t]
\begin{center}
\includegraphics[width=0.95\linewidth]{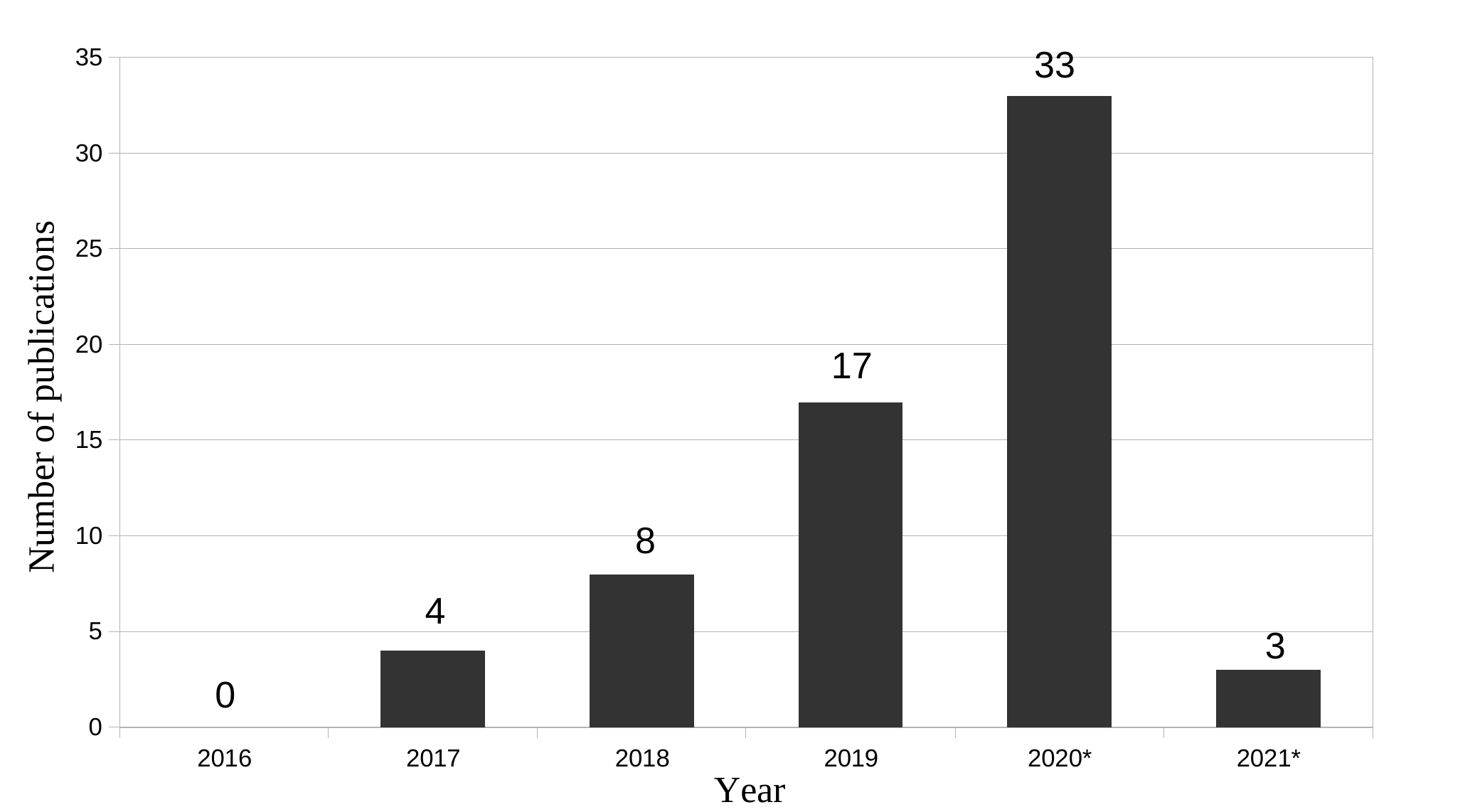}
\end{center}
\caption{Number of resulting publication per year by quering \textit{TITLE-ABS-KEY ( "blockchain"  AND  "tourism" )} in SCOPUS. For the years 2020 and 2021, the symbol * indicates the presence of incomplete data, updated to November 2020.}
\label{fig:process}
\end{figure}

%Tab. \ref{research} riporta una sintesi della letteratura scientifica scelta ed analizzata nel presente lavoro di tesi in merito all’applicazione della tecnologia blockchain nel turismo. La selezione di seguito riportata è avvenuta innanzitutto sulla base del contenuto descritto nell’abstract e ritenuto utile per lo svolgimento dell’argomentazione, ed in base alla reperibilità dei lavori, i quali sono stati ottenuti sia per mano degli autori che dai siti internet degli editori. Un altro criterio di scelta consiste nel grado di apprezzamento da parte dei cultori della materia: alcuni di questi articoli sono, per numero di citazioni, tra i principali riferimenti degli studiosi della blockchain applicata al turismo.

\subsection{Selected literature}

The selection process consisted of the following steps. From the list of  65 papers obtained from the above described query, the papers whose abstracts dealt with the application of blockchain technology to tourism as the main topic of the study were selected. Of these, the availability of the full text was verified (either by publishers’ websites  and by courtesy of the authors).  
Then, we have focused our attention on the scientific contributions of the last two years in relation to the general principles of the BOSE \cite{porru2017}. In particular, we chose the papers that discussed the opportunities offered by blockchain technology applied in the tourism sector. We have also included the papers that discussed the challenges that the tourism industry will have to face with the adoption of blockchain technology, both from a technological point of view and from an economic-organizational point of view. Tab. \ref{research} reports the summary of the scientific literature chosen and analysed in this work about the application of blockchain technology in tourism.

%Finally, of the available papers, both those with the greatest number of citations and those that offered the greatest contribution to the discussion on Blockchain-Oriented Software Engineering  were chosen.

%The selection was first made on the basis of the content described in the abstract and considered useful for the development of the argument, and on the basis of the availability of the works, which have been obtained either by authors and publishers’ websites. Another criterion of choice consists in the degree of appreciation by experts of the subject: some of these articles are, for number of citations, among the main references of scholars of the blockchain applied to tourism.

\begin{table*}[]
\caption{Summary of the research papers analyzed in this study}
\begin{tabular}{|l|l|l|l|}
\hline
\textbf{YEAR} & \textbf{TITLE}                                                                                                                  & \textbf{AUTHORS}                                                                                        & \textbf{MAIN CONTRIBUTIONS}                                                                                                                                                                                                                      \\ \hline
2018          &  Blockchain and Tourism: Three Research Propositions                                                                              & \cite{onder2018} Önder I. \& Treiblmaier H.                                                                             & \begin{tabular}[c]{@{}l@{}}Authors formulate three research hypotheses \\ about the impact of blockchain on tourist\\ industry.\end{tabular}                                                                                                \\ \hline
2019          & \begin{tabular}[c]{@{}l@{}}The Impact of Blockchain on the Tourism Industry: \\ A Theory-Based Research\end{tabular}             & \cite{treiblmaier2019} Treiblmaier H. \& Önder I.                                                                             & \begin{tabular}[c]{@{}l@{}}Understanding the potential impacts of \\ blockchain technology on tourist facilities and\\ how to cope with the changes induced by it.\end{tabular}                                                    \\ \hline
2019          & \begin{tabular}[c]{@{}l@{}}Is blockchain technology a watershed for tourism \\ development?\end{tabular}                         & \cite{kwok2019} Kwok A. O. J. \& Koh S. G. M.                                                                          & \begin{tabular}[c]{@{}l@{}}Opportunities and main challenges arising \\ from the use of blockchain technology for\\ small island economies.\end{tabular}                                                                               \\ \hline

2019          & \begin{tabular}[c]{@{}l@{}}Blockchain technology for smart city \\ and smart tourism: Latest trends and challenges\end{tabular}  & \begin{tabular}[c]{@{}l@{}} \cite{nam2019} Nam K., Dutt C. S., \\ Chathoth P. \& Khan M. S.\end{tabular}               & \begin{tabular}[c]{@{}l@{}}Latest blockchain technology trends and\\  challenges for smart cities and smart tourism \\ and their influences for the sector.\end{tabular}                                                                             \\ \hline
2019          & \begin{tabular}[c]{@{}l@{}}Assessment of blockchain applications in travel and \\ tourism industry\end{tabular}                  & \cite{ozdemir2020} Ozdemir A. I., Ar I. M. \& Erol I.                                                                     & \begin{tabular}[c]{@{}l@{}}A basic set of blockchain criteria is proposed \\ for the comparison of  various decentralized\\ applications.\end{tabular}                                                                                     \\ \hline
2019          & \begin{tabular}[c]{@{}l@{}}Blockchain Technology in Tourism: \\ Applications and Possibilities\end{tabular}                      & \cite{rejeb2019} Rejeb A. \& Rejeb K.                                                                                    & \begin{tabular}[c]{@{}l@{}}Potential effects and applications offered by \\ blockchain technology to the tourism\\  industry.\end{tabular}                                                                                            \\ \hline
2019          & \begin{tabular}[c]{@{}l@{}}A blockchain approach for the sustainability in \\ tourism management in the Sulcis area\end{tabular} & \begin{tabular}[c]{@{}l@{}} \cite{baralla2019} Baralla G., Pinna A., Tonelli R., \\ Marchesi M. \& Mannaro K.\end{tabular} & \begin{tabular}[c]{@{}l@{}}A blockchain platform is proposed for \\ traceability and certification of agri-food \\ products origin and for the promotion of \\ tourist activities of a destination.\end{tabular} \\ \hline
2020          & Blockchain and Tourism                                                                                                           & \cite{treiblmaier2020} Treiblmaier H.                                                                                         & \begin{tabular}[c]{@{}l@{}}Economic potential of blockchain technology\\ for the future of the tourism industry.\end{tabular}                                                                                                              \\ \hline
2021          & \begin{tabular}[c]{@{}l@{}}Applications of Blockchain and Smart Contract \\ for Sustainable Tourism Ecosystems\end{tabular}      & \cite{joo2021} Joo J., Park J. \& Han Y.                                                                              & \begin{tabular}[c]{@{}l@{}}Analysis of some cases of blockchain \\ technology application and smart contracts to \\ the tourism industry and identification of \\ opportunities to innovate existing companies.\end{tabular}         \\ \hline
\end{tabular}
\label{research}
\end{table*}

The analysis of the articles in Tab. \ref{research}, aims to be an opportunity to present the general framework of studies that address the relationship between blockchain technology and the tourism sector, to identify what may be its contribution to the development of this industry. In our survey, the important contribution of Önder and Treiblmaier stands out. They are among the most prolific authors dealing with the issue of the application of blockchain technology in tourism and enrich the ideas of reflection on this issue. In 2018, Önder and Treiblmaier \cite{onder2018} conducted a study in which they formulated three research hypotheses concerning the impact of the blockchain on the tourist industry, claiming that this will bring:
the activation of reliable systems for evaluating tourism products through the use of new forms of assessment and review technologies;
the creation of new consumer-to-consumer (C2C) markets through the adoption of cryptocurrencies;
disintermediation of the tourist industry.
Before buying a trip and its related services online, it is normal for any traveller to first read the reviews of those who have already used the same service, considering this information more up-to-date and reliable than that given directly by travel service providers. Indeed, it is well known that online reviews of tourism products have a significant influence on potential consumers. However, these reviews are often subject to falsification by industry players themselves (such as hotel or restaurant owners). They seek to orient customers to their advantage by creating a not entirely truthful evaluation of their products by posting fake reviews; this is complicated by the fact that anyone can publish travel reviews without being asked for proof of having completed the tourist experience. To overcome this situation, and thus have correct online reviews, through blockchain technology it is possible to create evaluation systems that provide individuals with traceable identities through unique private keys, in order to avoid distortions in reviews and make them more reliable. This would create a decentralized, impartial, and transparent system that guarantees authentic and reliable reviews to potential travellers, which once entered in the blockchain register would no longer be possible to modify or delete. In such a system, would still be guaranteed the privacy of users, who could also be more encouraged in making reviews with any financial rewards in the form of tokens and cryptocurrencies to be received as a reward.
Most tourism products often involve the transfer of money between partners located in different countries. This may require intervention by trusted intermediaries to conclude the transaction, with the consequence of having to pay additional large commissions; think for example of the 20\% charged by Booking to hoteliers, overpricing which inevitably affects final consumers. Using cryptocurrencies, however, it is possible to ensure an easy exchange of money without the help of third parties intermediaries and participate in the emergence of new forms of C2C markets of tourism products. Cryptocurrencies could therefore change the way tourists and operators exchange money and avoid the intervention of third parties (banks) in transactions, with consequent cost reductions. This is especially important where tourists' personal information, including financial data, cannot be entrusted to intermediaries. There are many systems that use blockchain to ensure high standards of security in transactions and at the same time greater protection of travellers’ information. One of these is TripEcoSys (10), which aims to be the largest decentralized tourism platform in the world based on the Ethereum blockchain. It is a system in which it is possible to find in a single space all the providers of the different travel services (flights, accommodation, excursions, etc.) that normally the visitor should book independently of each other, and where you can earn cryptocurrency rewards by sharing your own experiences. Furthermore, given the characteristics of cryptocurrencies, through blockchain networks it is also possible to avoid the problem of converting foreign currencies and, consequently, further limit the influence exerted by intermediaries, making transactions safer. If, on the one hand, tourists would no longer have to convert money when travelling, thus eliminating all the risks associated with the exchange of foreign currencies, on the other hand companies would be free to adjust their prices according exchange rate variability.
The advent of information technologies also in the tourism sector can therefore significantly reduce the intermediation chain towards the fruition of the tourist offer. The removal and replacement of intermediaries (also called disintermediation) is just one of the prerogatives of blockchain technology that in a short time will revolutionize the distribution and use of travel products and services, bringing significant benefits not only of an economic nature (cost savings). In general, it can be said that consumers will be more independent in organizing their travels. This disintermediation, however, does not only concern money transactions, but also directly affects the supply of tourism products. An initial form of disintermediation has already occurred with the advent of OTAs, but now also these could be replaced by blockchain-based, open source and decentralized travel platforms, such as WindingTree as discussed in the next section, which can eliminate the power exercised by intermediaries on the market. Another example is the Locktrip platform (see Table 2), which allows travellers to book hotels and other travel services without the intervention of any intermediary and, at the same, time provide feedback on the quality of the service. All this by using blockchain technology. From these simple examples, one can already understand the extent of the blockchain in the future of the tourist industry.
In another study, Treiblmaier \& Önder \cite{treiblmaier2019} also set out to analyse the potential impacts of blockchain technology on tourism facilities and how organizations could cope with the changes it induces. In general, by interviewing the managers of ten European Destination Management Organizations (DMOs), they highlighted that blockchain can be an important resource for many tourist organizations because it has the potential to change market structures. However, this is still a complex technology as it requires a certain investment by organizations, but they may not have the necessary financial resources and appropriate know-how to exploit it adequately. It follows, therefore, that this technology can be a new (important) resource only for those organizations that are in a position to exploit it for their own benefit, especially in a sector such as tourism. In 2020, Treiblmaier \cite{treiblmaier2020} summarizes and discusses the current state of the art, describing a list of twelve use cases of blockchain adoption in tourism, and the most relevant theoretical aspects, according to the academic literature. It focuses attention on the economic aspects of tourism disintermediation, highlighting this use case as looming in the tourism sector, and as a challenge that the tourism organization must be ready to face.

\begin{table*}[ht]
\caption{Active or proposed blockchain oriented software system for tourism industry}
\begin{tabular}{|l|l|l|l|l|l|}
\hline
\multicolumn{2}{|l|}{\textbf{NAME AND URL}}                                                              & \textbf{NATIONALITY} & \textbf{ACTIVITY STATUS} & \textbf{KIND OF APPLICATION}                                                                          & \textbf{BLOCKCHAIN} \\ \hline
\multicolumn{2}{|l|}{\begin{tabular}[c]{@{}l@{}}(1) WindingTree\\ https://windingtree.com/\end{tabular}}   & Switzerland             & Operating                  & Booking hotels and flights                                                                                   & Ethereum                       \\ \hline
\multicolumn{2}{|l|}{\begin{tabular}[c]{@{}l@{}}(2) LockTrip\\ https://locktrip.com/\end{tabular}}         & Bulgaria             & Operating                  & Booking hotels, holiday homes and flights                                                                     & Ethereum                       \\ \hline
\multicolumn{2}{|l|}{\begin{tabular}[c]{@{}l@{}}(3) FoodChain\\ https://food-chain.it/\end{tabular}}       & Italy               & Operating                  & Traceability of food products                                                                           & Quadrans                       \\ \hline
\multicolumn{2}{|l|}{\begin{tabular}[c]{@{}l@{}}(4) Bagtrax\\ https://bagtrax.eu/\end{tabular}}            & United Kingdom          & Operating                  & Baggage tracking                                                                                        & ND                             \\ \hline
\multicolumn{2}{|l|}{\begin{tabular}[c]{@{}l@{}}(5) Yookye\\ https://yookye.com/it\end{tabular}}           & Italy               & Operating                  & \begin{tabular}[c]{@{}l@{}}Organization of the holiday \\ (holiday home, services, experiences)\end{tabular} & Ethereum                       \\ \hline
\multicolumn{2}{|l|}{\begin{tabular}[c]{@{}l@{}}(6) DTCM Tourism 2.0\\  https://dubai10x.ae/\end{tabular}} & United Arab Emirates  & Announcement                   & \begin{tabular}[c]{@{}l@{}}Check occupancy status of hotel in Dubai\end{tabular}                & ND                             \\ \hline
\multicolumn{2}{|l|}{\begin{tabular}[c]{@{}l@{}}(7) WICKET\\ https://www.wicketevents.com/\end{tabular}}   & Italy               & Operating                  & Ticketing                                                                                                   & Ethereum                       \\ \hline
\multicolumn{2}{|l|}{\begin{tabular}[c]{@{}l@{}}(8) Trippki\\ https://trippki.com/\end{tabular}}           & United Kingdom          & Operating                  & Hotel booking                                                                                    & Ethereum                       \\ \hline
\multicolumn{2}{|l|}{\begin{tabular}[c]{@{}l@{}}(9) Travelchain\\ https://travelchain.io/\end{tabular}}    & Russia               & Prototype                  & Travel ecosystem                                                                                       & ND                             \\ \hline
\multicolumn{2}{|l|}{\begin{tabular}[c]{@{}l@{}}(10) TripEcoSys\\ https://www.tripecosys.com/\end{tabular}} & United Kingdom          & Prototype                  & Travel ecosystem                                                                                       & Ethereum                       \\ \hline
\multicolumn{2}{|l|}{\begin{tabular}[c]{@{}l@{}}(11) Sardcoin\\ https://www.sardcoin.eu/\end{tabular}}      & Italy               & Prototype                  & Smart coupon ecosystem                                                                                  & Hyperledger Fabric             \\ \hline
\end{tabular}
\label{projects}
\end{table*}

Kwok and Koh (2018)\cite{kwok2019} confirm the high potential of blockchain technology, analysing its opportunities in relation to the economies of small islands, which could exploit it to their advantage to compete with larger and often more renowned destinations. According to the two authors, in fact, blockchain technology represents a valid aid for such economies, strongly limited by their small size and their insularity, to implement tourism policies that encourage their economic growth. For example, Aruba is seeking to increase tourism revenue through the creation of a blockchain-based platform for travel bookings, while the Caribbean islands are promoting the adoption of their own regional cryptocurrency. In particular, the effects resulting from the application of the blockchain to small island tourism can be traced back to four categories that can create beneficial effects for both tourists and destinations. In summary, these are: the general offer of a better tourist experience to visitors; greater speed and security in transactions with foreign countries thanks to the use of cryptocurrencies; diversification of the financial offer and use of state-owned cryptocurrencies; finally, a reduction in costs for host destinations. All this, however, can only be possible if foreign tourists are able to understand it and if local operators are able to accept it, thus avoiding limiting its use to a restricted group of experienced users.

Nam et al. (2019)\cite{nam2019} analysed the latest trends and challenges regarding blockchain technology for smart cities and smart tourism, focusing their attention on the comparison of thirteen decentralized applications (Dapps) of the tourism industry. In particular, scholars have formulated some research proposals about its evolution and influence in the industry, coming to the conclusion that the adoption of blockchain technology, and especially Dapps, will lead in the future to the creation of new business models and new market structures. Although a limited number of solutions have been analysed, especially when compared to the totality of existing blockchain solutions, their analysis revealed three characteristics common to all Dapps: cost reduction, increasing adoption of cryptocurrencies and development of new all-encompassing ecosystems. More precisely, a market will be created with strong competition between new online travel platforms based on blockchain and those already existing on the market, that to survive will be forced to change their business models and adapt to new emerging trends. Furthermore, since it will not be easy to ensure the adoption of blockchain technology by travellers and other stakeholders, some operators may also provide for the granting of incentives for the use of cryptocurrencies (the greater the incentives, the faster the adoption of blockchain technology). This will therefore lead to the identification of some "dominant players", that is those platforms with the largest number of users, which will impose themselves on the market.

%Anche Ozdemir et al. (2019) hanno effettuato un confronto tra diverse DApps, fornendo alcuni esempi pratici di come queste sono utilizzate nel turismo. In particolare, essi si sono concentrati, però, sulla proposta di un set di criteri di base della tecnologia blockchain che può rivelarsi utile per valutare le applicazioni decentralizzate. A tale scopo, secondo gli autori occorre tener conto del modello di governance della blockchain, della piattaforma su cui essa è implementata, del tipo di consenso, dell’utilizzo di criptovalute, dell’uso di smart contracts ed infine dei token. Essi sostengono, infatti, che la comprensione di questi elementi è il punto centrale per comprendere e migliorare qualsiasi applicazione blockchain. Ad esempio, l’individuazione del modello di governance applicato alla blockchain è di fondamentale importanza per determinare le sue caratteristiche, così come la tipologia di piattaforma su cui è implementata può influire sulle relative prestazioni. Tuttavia, le loro considerazioni necessitano di ulteriori sviluppi, soprattutto in considerazione del campione di DApps analizzato: alcune di queste applicazioni, infatti, sembrano non più esistenti.
Ozdemir et al. (2019)\cite{ozdemir2020} also made a comparison between several DApps, providing some practical examples of how these are used in tourism. In particular, they have focused, however, on proposing a basic set of blockchain technology criteria that may prove useful for evaluating decentralized applications. For this purpose, according to the authors, it is necessary to consider the governance model of the blockchain, the platform on which it is implemented, the type of consent, the use of cryptocurrencies\cite{}, the use of smart contracts and finally tokens. They argue that understanding these elements is the key to understand and improve any blockchain application. For example, the identification of the governance model applied to the blockchain is of fundamental importance to determine its characteristics, as well as the type of platform on which it is implemented can affect its performance\cite{destefanis2018}. However, their considerations need further development, especially considering the sample of DApps analysed: some of these applications, in fact, seem to no longer exist.

%Lo studio condotto da Baralla et al. (2019), invece, è finalizzato a proporre una piattaforma blockchain per la tracciabilità e la certificazione di provenienza dei prodotti agroalimentari sardi e per la promozione delle attività turistiche del Sulcis, in Sardegna, secondo un’ottica di sostenibilità. Grazie alle caratteristiche della blockchain, infatti, è possibile garantire informazioni sicure e trasparenti e far in modo che arrivino a tutti gli stakeholder coinvolti. Seguendo tale logica, il sistema proposto da Baralla et al. si avvale di una serie di smart contracts per garantire la tracciabilità e la provenienza dei prodotti, dando al turista/consumatore la possibilità di verificarne l’autenticità e di fornire propri feedback attraverso la pubblicazione di messaggi, foto e video. In tal modo, gli operatori turistici potranno contestualmente migliorare i propri servizi ed arricchire l’offerta turistica del territorio. Nonostante il sistema proposto sia ancora in divenire e limitato ad un territorio circoscritto, esso risulta comunque molto promettente anche per la futura gestione turistica di territori ben più ampi.
The study conducted by Baralla et al. (2019)\cite{baralla2019}, however, aims at proposing a blockchain platform for traceability and certification of origin of Sardinian agri-food products and for the promotion of tourist activities in Sulcis area, in Sardinia, with a view to a sustainability. Thanks to the blockchain characteristics, in fact, it is possible to guarantee secure and transparent information and ensure that it reaches all the stakeholders involved. Following this logic, the system proposed by Baralla et al.  makes use of a series of smart contracts to ensure traceability and provenance of products, giving tourist/consumer the opportunity to verify their authenticity and provide their own feedback by posting messages, photos, and videos. In this way, tour operators can simultaneously improve their services and enrich the tourist offer of the area. Although the proposed system is still in progress and limited to a finite territory, it is still very promising also for the future tourist management of much larger territories.

%Joo et al. (2021), infine, hanno concentrato la loro attenzione sullo studio della tecnologia blockchain e dei relativi smart contracts nel settore del turismo sostenibile e sull’individuazione delle migliori opportunità per innovare le imprese esistenti, fornendo alcuni esempi applicativi. Anch’essi confermano quanto evidenziato già da altri studiosi circa l’utilità della blockchain per migliorare la trasparenza e la sicurezza delle operazioni di viaggio, la fiducia degli utenti e la riduzione dei costi di transazione. Gli esempi proposti, inoltre, possono servire da stimolo per gli operatori turistici che vogliono rendere i loro business più innovativi.
Finally, Joo et al. (2021)\cite{joo2021} focused their attention on the study of blockchain technology and related smart contracts in the sustainable tourism sector and on identifying the best opportunities to innovate existing companies, providing some application examples. They also confirm the evidence already highlighted by other scholars about the usefulness of the blockchain to improve the transparency and safety of travel operations, users’ trust, and the reduction of transaction costs. The proposed examples can also help as a stimulus for tour operators who want to make their businesses more innovative.

\section{Projects and applications of blockchain technology in tourism}

\subsection{Relevant Business initiatives and Research Projects }

In Tab. \ref{projects} we report eleven blockchain oriented software that are used in the tourism industry. The list presented is the result of two considerations: there are some software that are cited as relevant examples in the reference literature, and others that have emerged from a specific research on the main uses of blockchain technology in the tourism sector. In addition to the name and their URL, for each of them are indicated: the nationality, the activity status (distinguishing the operating software, from the announcements and the prototypes), their main functionality and, where available, the type of blockchain technology used.

\subsection{Fields of application of blockchain technology}
%Nonostante in precedenza vi sia già fatto riferimento ad alcune di esse, nell’industria del turismo e dell’ospitalità possono esserci molteplici potenziali applicazioni della tecnologia blockchain, alcune delle quali possono anche integrarsi tra loro.
%Di seguito si riportano sinteticamente alcune delle principali applicazioni della blockchain nel turismo e di quelle che potrebbero essere realizzate per aggiungere valore al settore.

Although some of them have already been mentioned previously, in the tourism and hospitality industry there may be many potential applications of blockchain technology, some of which may also integrate with each other.
Some of the main applications of blockchain in tourism and those that could be implemented to add value to the sector are summarized below.

\subsubsection{Inventory management} %La blockchain può essere utilizzata in un sistema di gestione delle scorte che nel settore dell’ospitalità può riferirsi al numero di camere disponibili negli alberghi o al numero di posti disponibili negli aerei. Nello specifico, può fornire informazioni sulle disponibilità e sul tasso di copertura condividendole con i vari stakeholder interessati, quindi sostituendosi, di fatto, ad eventuali Property Management System (PMS) e CRS, con la conseguente rimozione degli intermediari e delle relative spese. Esemplare in tal senso è la già citata piattaforma svizzera WindingTree, che consente agli albergatori e alle compagnie aeree di elencare le proprie disponibilità, e ai turisti di prenotarli. Un altro esempio è dato da “BedSwap”, un progetto al quale sta lavorando il gruppo tedesco TUI (Tourism Union International) per una gestione efficace delle camere d’albergo dei propri partner nei mercati serviti dalla compagnia stessa. Il Dipartimento del Turismo e del Commercio di Dubai (DTCM), invece, nel 2018 ha lanciato “Tourism 2.0”, con l’obiettivo di far diventare Dubai la prima destinazione al mondo per viaggi ed eventi globali entro il 2020. Si tratta di un sistema che consente di verificare l’occupazione degli hotel e le tariffe delle camere, permettendo agli altri operatori turistici di predisporre le proprie offerte in modo più efficace per i clienti.
Blockchain technology can be used in an inventory management system which in the hospitality sector can refer to the number of rooms available in hotels or the number of seats available on airplanes. Specifically, it can provide information on availability and coverage rate by sharing it with the various interested stakeholders, thus replacing any Property Management System (PMS) and CRS, with the consequent removal of intermediaries and related expenses. An example of this is the Swiss platform WindingTree (1) , which allows hoteliers and airlines to list their availabilities, and tourists to book them. Another example is given by "BedSwap", a project on which the German group TUI (Tourism Union International) is working for an effective management of the hotel rooms of its partners in the markets served by the company itself. The Dubai Department of Tourism and Commerce (6), on the other hand, in 2018 launched "Tourism 2.0", with the aim of making Dubai the world’s first destination for global travel and events by 2020. It is a system that allows you to check hotel occupancy and room rates, allowing other tour operators to prepare their offers more effectively for customers.

\subsubsection{Traceability of food products} %in un contesto turistico sempre più globalizzato è importante dotare le destinazioni di strumenti che garantiscano la tracciabilità nell’approvvigionamento dei prodotti della catena alimentare, soprattutto in quei settori del turismo enogastronomico il cui vantaggio competitivo deriva dall’impiego di prodotti biologici, locali e sostenibili. La blockchain può essere utilizzata per creare un sistema che consenta alle persone di accedere ai dati sulla provenienza dei prodotti alimentari e quindi ricostruire il percorso dei prodotti dal campo alla tavola. In termini pratici ciò potrebbe essere fatto scansionando un QR Code o direttamente i codici a barre dei prodotti attraverso dispositivi registrati nella blockchain. Ad esempio, Foodchain è un sistema italiano che, avvalendosi anche dell’Internet of Things (IoT) traccia i prodotti alimentari dall’origine fino al consumatore finale utilizzando la tecnologia blockchain. Le informazioni rese pubbliche dalle aziende partecipanti sono accessibili ai consumatori finali attraverso un QR Code applicato sul packaging dei prodotti, il che contribuisce ad incrementare la loro fiducia nei confronti delle aziende vista la possibilità di controllare tutte le informazioni sul prodotto. 
In an increasingly globalized tourism context, it is important to provide destinations with tools to ensure traceability in the supply of food chain products, especially in those sectors of food and wine tourism whose competitive advantage derives from the use of organic, local, and sustainable products. Here blockchain can be used to create a system that allows people to access data on the origin of food products and then reconstruct the path of products from the field to the table. In practical, this could be done by scanning a QR Code or directly barcodes of products through devices registered in the blockchain. For example, Foodchain (3) is an Italian system that, also using the Internet of Things (IoT), traces food products from origin to the final consumer using blockchain technology. The information made public by the participating companies is accessible to end consumers through a QR Code applied on the packaging of products. This helps to increase their confidence in companies given the possibility to check all the information on the product.

\subsubsection{Baggage tracking} %Nonostante i progressi fatti dalle compagnie aeree nella gestione dei bagagli, molti viaggiatori continuano a riscontrare problemi di bagagli smarriti, con conseguenti perdite di tempo e denaro tanto per i passeggeri quanto per le compagnie stesse. La tecnologia blockchain si propone come soluzione in tal senso poiché consente il tracciamento dei bagagli, che possono essere monitorati in vari punti essenziali, attraverso l’inserimento automatico dei dati rilevati in un registro pubblico, velocizzando così anche il check-in e riducendo di fatto i tempi di attesa in aeroporto. Ad esempio, Bagtrax utilizza un sensore che, una volta attaccato alle valigie, consente di localizzare i bagagli durante i trasferimenti e di chiedere un risarcimento immediato in caso di smarrimento. Questo sistema si avvale della tecnologia blockchain per rendere le sequenze di tracciamento protette e garantire la sicurezza dei dati di tutti gli attori coinvolti nel servizio: i passeggeri, le compagnie aeree, i gestori aeroportuali e le compagnie assicurative. 
Despite the progress made by airlines in handling luggage, many travellers continue to experience problems with lost luggage, which leads to a loss of time and money for both passengers and airlines themselves. Blockchain technology is proposed as a solution in this sense since it allows the tracking of baggage, which can be monitored at various essential points, through the automatic entry of the data collected in a public register, thus also speeding up check-in and effectively reducing waiting times at the airport. For example, Bagtrax (4) uses a sensor which, once attached to suitcases, allows you to locate your luggage during transfers and to claim immediate compensation in case of loss. This system uses blockchain technology to make tracking sequences protected and guarantee the data security of all the actors involved in the service: passengers, airlines, airport managers and insurance companies.

\subsubsection{Reservations and ticketing} %In tal caso l’utilità della blockchain risiede in possibili utilizzi legati all’effettuazione di prenotazioni, per l’emissione di biglietti e per contrastare il mercato nero, ad esempio mediante la creazione di protocolli standard che consentano agli acquirenti di utilizzare i loro portafogli elettronici per dimostrare la proprietà del biglietto. Rientra in quest’ambito WICKET, un’app lanciata da una startup italiana che utilizza un particolare protocollo basato sulla tecnologia blockchain per digitalizzare i biglietti e porre un limite al cosiddetto fenomeno del bagarinaggio. Essa utilizza il protocollo GET (Guaranteed Entrance Token), già impiegato in Olanda e Singapore per la vendita di biglietti di eventi sportivi o di altre manifestazioni quali concerti, fiere e congressi. Quest’anno però, a seguito della pandemia che ha colpito l’intero pianeta, è stata proposta agli stabilimenti balneari per consentire ai bagnanti di prenotare il proprio posto in spiaggia e pagare direttamente online. Il sistema associa il biglietto al numero di telefono dell'acquirente, generando un Qr code univoco in un apposito wallet, e lo renderà fruibile all'utente tramite app per consentire a quest'ultimo di accedere alla struttura.
In this case, the blockchain’s usefulness lies in possible uses related to making reservations, for the issuance of tickets and to contrast the black market, for example by creating standard protocols allowing buyers to use their electronic wallets to prove ownership of the ticket. This includes WICKET (7), an app launched by an Italian startup that uses a special protocol based on blockchain technology to digitize tickets and limit the phenomenon of speculation. It uses the GET (Guaranteed Entrance Token) protocol, already used in the Netherlands and Singapore for the sale of tickets for sporting events or other events such as concerts, fairs and congresses. This year, however, following the pandemic that has affected the entire planet, was proposed to bathing establishments to allow bathers to book their place on the beach and pay directly online. The system associates the ticket with the buyer's phone number, generating a unique QR code in a special wallet, and will make it accessible to the user via an app to allow the latter to access the facility.

\subsubsection{Travellers loyalty} %Gli operatori del settore turistico possono avvantaggiarsi dalla creazione di appositi programmi fedeltà che rilasciano token come ricompensa ai viaggiatori. È stato dimostrato, infatti, che le aziende che fanno ricorso a sistemi di questo tipo riescono ad ottenere un vantaggio competitivo rispetto ai concorrenti, a raggiungere nuovi potenziali clienti e a migliorare la percezione del proprio marchio ai loro occhi, rafforzando quindi di il legame tra viaggiatori e destinazioni. La startup Loyyal ha lanciato una piattaforma basata sulla blockchain attraverso cui le aziende possono gestire propri programmi fedeltà offrendo sistemi di ricompense di vario tipo. Il Gruppo arabo Jumeirah, in collaborazione con Dubai Holding, utilizza proprio questa piattaforma per migliorare l'efficienza dei suoi programmi di fidelizzazione. Trippki, invece, ha ideato un programma di fidelizzazione per consentire a clienti e aziende del settore turistico di entrare direttamente in contatto tra loro, promuovendo così la disintermediazione del settore. Nello specifico, ai clienti vengono assegnati alcuni token per soggiornare in un determinato hotel che vengono registrati nella blockchain senza scadenza, garantendo quindi la possibilità di riscattarli in qualsiasi momento.
Travel industry operators can benefit from the creation of dedicated loyalty programs that issue tokens as rewards to travellers. In fact, it has been demonstrated that companies that use such systems are able to obtain a competitive advantage over their competitors, reach new potential customers and improve the perception of the brand in their eyes, thus strengthening the link between travellers and destinations. The startup Loyyal has launched a blockchain-based platform through which companies can manage their loyalty programs by offering various types of reward systems. The Arab Group Jumeirah, in collaboration with Dubai Holding, uses this platform to improve the efficiency of its loyalty programs. Trippki (8) , on the other hand, has devised a loyalty program to allow customers and companies in the tourism sector to enter directly into contact with each other, thus promoting the disintermediation of the sector. Specifically, customers are given some tokens (that are registered in the blockchain without expiration date) to stay in a certain hotel, thus ensuring the possibility to redeem them at any time.

\subsubsection{Identity, credential management and privacy} %al fine di migliorare la sicurezza del settore e proteggere la privacy dei viaggiatori, problemi di cui si è già parlato in precedenza, si potrebbe pensare alla definizione di un’identità di viaggiatore (digitale) globale attraverso un sistema che consenta di determinare in maniera inequivocabile l’identità di una persona, nel contempo trovando una possibile soluzione al problema del furto d’identità cui spesso vanno incontro i turisti. Ad esempio, sarà possibile aggiungere alle varie informazioni registrate anche quelle biometriche (impronte digitali, riconoscimento facciale), semplificando così anche il lavoro degli hotel, che si dovranno limitare a registrare nella blockchain solo le date di arrivo e partenza degli ospiti, senza doverli più segnalare alla polizia o alle altre autorità. Da questo punto di vista, la società di telecomunicazioni del trasporto aereo SITA (Société Internationale de Télécommunications Aéronautiques) sta studiando come l'utilizzo di passaporti virtuali o digitali  possa ridurre i controlli dei documenti durante i viaggi dei passeggeri, eliminando così la necessità di possedere diversi documenti di viaggio. Attraverso un unico token contenente dati biometrici ed altre informazioni personali, ed archiviato su dispositivi mobili o indossabili, i viaggiatori possono infatti essere rapidamente identificati una sola volta da qualsiasi autorità.
In order to improve the sector safety and protect the privacy of travellers, we could think of the definition of a (digital) global traveller identity through a system that allows to determine unequivocally the identity of a person. At the same time, it is also possible finding a solution to the problem of identity theft that tourists often encounter. For example, it will be possible to add biometric information (especially fingerprints and facial recognition) to the other registered information thus also simplifying the work of hotels, that will have only to record in the blockchain the arrival and departure dates of guests, without having to report them to the police or other authorities. From this point of view, the telecommunications company of air transport SITA (Société Internationale de Télécommunications Aéronautiques) is studying how the use of virtual or digital passports can reduce document checks during passenger journeys, thus eliminating the need to possess various travel documents. Through a single token containing biometric data and other personal information, and stored on mobile or wearable devices, travellers can in fact be quickly identified only once by any authority.

\subsection{Technology}
Apart from their functionality in the tourism sector, there are two main blockchain technologies adopted: smart contracts and DApps.

\subsubsection{Smart contracts} %la loro programmabilità e l’esecuzione automatizzata indipendentemente da interferenze umane offre numerose potenzialità per l’industria del turismo. Ad esempio, questi consentono: l’attivazione di sistemi di pagamento immediato nelle transazioni, facilitando così la collaborazione tra hotel e agenzie di viaggio; l’assegnazione delle camere agli ospiti tramite chiavi digitali sulla blockchain; nelle compagnie aeree possono facilitare l’assicurazione del volo pagando automaticamente quanto concordato in caso di ritardo o cancellazione. La maggior parte delle applicazioni citate come esempio in questa sezione utilizzano, appunto, differenti smart contracts nei loro sistemi.
Their programmability and automated execution independently of human interference offers many potentials for the tourism industry. For example, these allow: the activation of immediate payment systems in transactions, thus facilitating collaboration between hotels and travel agencies; the allocation of rooms to guests via digital keys on the blockchain; airlines can facilitate flight insurance by paying automatically as agreed in case of delay or cancellation. Most of the applications mentioned as an example in this section use, in fact, different smart contracts in their systems.

\subsubsection{dApps for smart tourism} %sono in corso di definizione progetti per sistemi di recensioni online, la pianificazione dei viaggi, la comunicazione diretta con i proprietari di immobili, il marketing personalizzato. Ad esempio, la startup innovativa italiana Yookye aspira ad offrire ai turisti proposte di viaggio create su misura da esperti locali in base alle esigenze e aspirazioni degli utenti. Sulla base delle preferenze riscontrate, gli esperti locali, avvalendosi della propria conoscenza territoriale e dell’intelligenza artificiale, creano poi alcune proposte di viaggio tra cui l’utente potrà scegliere. Il tutto viene reso più affidabile grazie alla tecnologia Blockchain, che garantisce transazioni in totale sicurezza. Gli utenti godono in tal modo di un servizio di prenotazione di immobili a breve termine sicuro e affidabile, che affianca ai tradizionali metodi di pagamento le criptovalute più famose e il token YOOK della stessa piattaforma.
Are in the process of defining projects for online review systems, travel planning, direct communication with property owners, personalized marketing. For example, the innovative Italian startup Yookye (5) aims to offer tourists travel proposals tailored by local experts on the needs and aspirations of users. On the basis of the preferences found, the local experts, making use of their territorial knowledge and artificial intelligence, then create some travel proposals from which the user can choose. Everything is made more reliable thanks to the blockchain technology, which guarantees transactions in total security. In this way, users enjoy a secure and reliable short-term property booking service, which combines traditional payment methods with the most famous cryptocurrencies and the YOOK token of the same platform.

\subsubsection{Breaking technological barrier}
%One of the problems arising from the use of new technologies in tourism is the lack of public confidence. Blockchain technology would seem to provide a solution, as the strengthening of trust represents one of the potential effects deriving from its application to tourism, especially in a period of great uncertainty like the one we are currently experiencing. The formation of trust in the hospitality industry, in fact, is still a little-known aspect, since it depends on the subjectivity of individual tourists, or rather on the risk that they are willing to accept in travel experiences. In the specific case of blockchain technology, the protocols on which it is based are structured in such a way as to ensure greater involvement by the various tourism stakeholders and thus improve their experience in the sector. 
Through the promotion of transparent transactions, blockchain technology guarantees a higher level of trust and security in online travel platforms, since all data will only be upgradable through a consensual agreement between all participants in the network. From this, it is clear that blockchain can be a valid tool for neutrality and objectivity in travel information systems. In this kind of system, in fact, customers/tourists feel more comfortable in sharing their travel experiences in a more open and sincere way. Examples include TravelChain (9) and WindingTree(1). The first is a blockchain-based travel business that rewards travellers for the transparent sharing of information regarding their experiences; the other, instead, is a decentralized and open-source platform that aims to find a solution to several problems in the travel industry (such as high commissions, obsolete technology, high entry barriers and lack of innovation), all linked to a high degree of centralization. This is a system in which transaction data is grouped into blocks and replicated among all participants to ensure greater transparency and control of the travel package. In this way, customers are able to ascertain the actual value of a tourism product and gain more control and power over the planning of their travel experience.

\section{Discussion}
The analysis carried out revealed two main effects that blockchain technology can bring to the tourism sector.

\subsubsection{Disintermediation} %della progressiva rimozione degli intermediari si è già parlato ampiamente. Quello che non si è detto, però è che, nello specifico, nel settore turistico la blockchain può portare alla sostituzione dei sistemi di distribuzione globale (GDS) che consentono transazioni tra diversi fornitori di servizi (ad esempio hotel, compagnie aeree, agenzie di viaggi, ecc.) e delle OTA attraverso sistemi che consentono comunicazioni e transazioni peer-to-peer.

The progressive removal of intermediaries has already been widely discussed. 
Today, travel agencies add a level of intermediation to the tourism chain, capable of generating end-to-end trust, that is, between the consumer (the tourist) and the tourist operator (i.e the DMO). Thanks to the presence of a trustless mechanism, blockchain technology makes the role of travel agencies accessory, that will have to review their role in the tourism system (as experts in the sector or as facilitators). This is a challenging opportunity for traditional travel agencies (which will be able to count on direct human relations).
What has not been said, however, is that, specifically in the tourism sector, blockchain can lead to the replacement of Global Distribution Systems (GDS) that allow transactions between different service providers (for example hotels, airlines, travel agencies , etc.) and OTAs through systems that allow peer-to-peer communications and transactions.

\subsubsection{Coordination and coopetition} %Grazie alla blockchain è possibile realizzare sistemi che vedono l’aggregazione di diversi prodotti e servizi di viaggio in modo tale da ridurre così eventuali inefficienze e realizzare un maggior coordinamento tra le varie proposte. La già citata piattaforma Yookye, ad esempio, raggruppa in un unico sistema numerosi servizi e utilities per il turista, semplificando l’organizzazione della sua vacanza. I suoi utenti, quindi, non avranno più bisogno di molteplici App, ma possono trovare e prenotare alloggi, tour ed altri servizi in un’unica App. Allo stesso modo Sardcoin, progetto dell’Università di Cagliari, offre una piattaforma per la vendita di servizi turistici sardi basata sulla tecnologia blockchain. Le imprese aderenti potranno usare l’infrastruttura per inserire servizi, integrare le proprie offerte o per impostare strategie di promozione turistica integrate con altre infrastrutture esistenti.
Thanks to the blockchain technology it is possible to create systems that see the aggregation of different travel products and services in such a way as to reduce any inefficiencies and achieve greater coordination between the various proposals. The aforementioned Yookye platform (5), for example, brings together in a single system numerous services and utilities for tourists, simplifying the organization of their vacation. Its users, therefore, will no longer need multiple Apps, but can find and book accommodation, tours and other services in a single App. Similarly, Sardcoin (11), a project of the University of Cagliari, offers a platform for the sale of Sardinian tourist services based on blockchain technology. That typology of systems implements the concept of \textit{coopetition}, that is the constructive and collaborative competition between companies from which both benefit: participating companies will be able to use the infrastructure to include services, integrate their offers or to set up tourism promotion strategies with other existing infrastructures.

\section{Conclusions}
On the basis of what has been said so far, it can be also said that tourism is one of the most promising sectors for the development of blockchain technology.

According to the results reported in this paper, we are able to answer the research questions which guided us in this investigation. In particular, for what concerns the first research question we can say that even if the number of the published research paper is still limited (to a total of 65 in the SCOPUS database), the number of published paper per year doubles every year. We examined the content of nine selected paper from which we can analyze the directions of scientific research in this domain. The second research question allowed us to explore the blockchain-oriented software projects for tourism. We reviewed 11 software projects, many of them operational, providing a classification of their core functionality and the related technological aspects.

%RQ1: to what extent does the scientific community address the problems relating to the use of blockchain technology in tourism? and

%RQ2: what is the state of the art of the practical use of blockchain technology in tourism and DMOs?

However, regardless of what the scope of the investigation, some other important aspects need to be taken into account in further research.
First of all, it is necessary to consider that the blockchain is just a collective term that synthesizes a series of different tools; in fact, there are numerous blockchain protocols that determine as many implementations. Secondly, it must be borne in mind that it is a constantly developing technology which involves from the simplest protocols that form the basis of it, up to real new applications. This progress is aimed above all at reducing the complexity of the technology regarding the difficulties of understanding (and consequently also of application). Finally, it must be said that, despite continuous progress, blockchain technology cannot be tied to a single use but lends itself to numerous purposes, even if sometimes conflicting with each other. In addition, it can also be combined with other technologies, such as the Internet of Things (IoT) or artificial intelligence, which integrate with each other to achieve increasingly innovative applications. This is confirmed by the blockchain technology application cases reported in this paper. This is the case of FoodChain (3) and Bagtrax (4), which integrate blockchain with the IoT to achieve their goals, or even of the startup Yookye (5), which combines blockchain technology and artificial intelligence tools.

%[Future works]
In this work we highlighted the most relevant issues of the blockchain technology applied to the tourism industry. However, it has emerged that the uses of blockchain technology are limited to initiatives by individual companies, so we recommend destination managers to deepen and encourage its application in the tourist management of the territories. Researchers could conduct empirical studies to assess the design of comprehensive systems that help managers in promoting this innovative tool.	
\bibliographystyle{plain}

\end{document}